# Photonics-assisted microwave pulse detection and frequency measurement based on pulse replication and frequency-to-time mapping

Pengcheng Zuo[a,b], Dong Ma[a,b], Qingbo Liu[c], Lizhong Jiang[c,d], and Yang Chen[a,b]*

[a]*Shanghai Key Laboratory of Multidimensional Information Processing, East China Normal University, Shanghai 200241, China*
[b]*Engineering Center of SHMEC for Space Information and GNSS, East China Normal University, Shanghai 200241, China*
[c]*Shanghai Radio Equipment Research Institute, Shanghai 201109, China*
[d] *School of Artificial Intelligence and Automation, Huazhong University of Science and Technology, Wuhan 430074, China,*

**Abstract**

A photonics-assisted microwave pulse detection and frequency measurement scheme is proposed. The unknown microwave pulse is converted to the optical domain and then injected into a fiber loop for pulse replication, which makes it easier to identify the microwave pulse with large pulse repetition interval (PRI), whereas stimulated Brillouin scattering-based frequency-to-time mapping (FTTM) is utilized to measure the carrier frequency of the microwave pulse. A sweep optical carrier is generated and modulated by the unknown microwave pulse and a continuous-wave single-frequency reference, generating two different frequency sweep optical signals, which are combined and used as the probe wave to detect a fixed Brillouin gain spectrum. When the optical signal is detected in a photodetector, FTTM is realized and the frequency of the microwave pulse can be determined. An experiment is performed. For a fiber loop containing a 210-m fiber, pulse replication and FTTM of the pulses with a PRI of 20 μs and pulse width of 1.20, 1.00, 0.85, and 0.65 μs are realized. Under a certain sweep frequency chirp rate of 0.978 THz/s, the measurement errors are below ±12 and ±5 MHz by using one pair of pulses and multiple pairs of pulses, respectively. The influence of the sweep frequency chirp rate and pulse width on the measurement error has also been studied. To a certain extent, the faster the frequency sweep, the greater the frequency measurement error. For a specific sweep frequency chirp rate, the measurement error is almost unaffected by the pulse width to be measured.

## 1. Introduction

Microwave pulses are widely utilized in pulse radar [1] and electronic warfare [2]. With the development of electronic warfare, directed energy weapons and electromagnetic pulse weapons are used to attack enemy personnel, facilities and equipment, thereby reducing or destroying the enemy's combat effectiveness. For example, in electromagnetic weapon attacks, ultra-short microwave pulses with large pulse repetition interval (PRI) and ultra-high peak power are used to destroy the receivers. Therefore, it is significantly important for the countermeasure system to quickly detect the microwave pulses and identify their parameters, which enables the receiving system to obtain the parameters of the attacks, avoid receiver damage, and continue to work through fast system agility. Therefore, it is highly desirable that the pulse parameters can be obtained in a single pulse cycle, which will avoid the damage of the receiver to the greatest extent. Pulse replication can provide a good solution to capture and identify the ultra-short microwave pulses. However, pulse replication in the electrical domain suffers from a large loss and is susceptible to electromagnetic interference (EMI).

* Corresponding author. Tel.: +86-21-33503290.
  *E-mail address:* ychen@ce.ecnu.edu.cn.

Recently, a pulse replication system based on an active fiber loop has been demonstrated [3], which can be utilized to detect the short microwave pulse. It is important to detect whether there is an ultra-short and high-power microwave pulse, which can help to turn off the receiver in time to avoid damage by the attack. However, it is more important to acquire the frequency of the microwave pulse, because it makes it possible for us to avoid the attack of the microwave pulse and ensure the normal operation of the system through frequency agility. The frequency measurement method based on conventional electrical means is susceptible to EMI and suffers from difficulties in achieving large bandwidths due to the well-known electronic bottleneck [4], [5].

Microwave photonics (MWP) focuses on the generation, processing, control, and measurement of microwave signals, taking the advantages of large bandwidth, high frequency, good tunability, and immunity to EMI, offered by modern optics [6], [7]. Numerous photonics-based methods have been reported for the measurement of microwave parameters during the past few decades. Microwave frequency measurement is one of them, which can be divided into three categories, i.e., frequency-to-time mapping (FTTM) [8-10], frequency-to-power mapping [11-13], and frequency-to-space mapping [14-16]. Stimulated Brillouin scattering (SBS) is a typical nonlinear effect caused by the acousto-optic interaction in optical fibers. The Brillouin gain spectrum produced by the SBS effect has a narrow bandwidth, good wavelength tunability, and low threshold, which has attracted great attention in many fields [17-20], one of which is microwave frequency measurements. Recently, we have proposed a multiple microwave frequency measurement approach based on SBS and FTTM [21], in which the measurement accuracy is better than ±1 MHz by introducing a two-step accuracy improvement. To improve the microwave frequency measurement resolution, we have proposed a multiple RF frequency measurement method based on the reduced SBS gain spectrum [22], in which a resolution of less than 10 MHz is obtained. However, nearly all the photonics-assisted frequency measurement approaches are designed for continuous-wave (CW) microwave signals. As discussed above, in modern electronic warfare systems, it is highly desirable to measure the frequency of a short microwave pulse with a large PRI in time.

In this paper, we propose and experimentally demonstrate a photonics-assisted microwave pulse detection and frequency measurement scheme based on pulse replication and SBS-based FTTM. To capture and identify a microwave pulse with a large PRI in time, the unknown electrical pulse is converted into the optical domain and then injected into a fiber loop for pulse replication. To measure the carrier frequency of the microwave pulse, a periodic sweep optical signal is generated through carrier-suppressed lower single-sideband (CS-LSSB) modulation by an electrical sweep signal and then used as a new optical carrier. A fixed-frequency CW reference and the unknown electrical pulse are carrier-suppressed double-sideband (CS-DSB) modulated onto the frequency sweep optical carrier to generate two different frequency sweep signals. In this way, a fixed Brillouin gain spectrum can be swept by the two different frequency sweep optical signals to realize the FTTM. Accordingly, two kinds of low-frequency electrical pulses are generated at different specific times in some measurable period. By using the time difference between two pulses corresponding to the CW reference and microwave pulse, the frequency of the microwave pulse can be obtained. To the best of our knowledge, this is the first time that a pulsed microwave signal is measured with the help of pulse replication and SBS-based FTTM. The replication of pulses in the optical domain provides the possibility to detect microwave pulses with a large PRI in time, while the position of multiple replicated pulses can be used for pulse frequency measurement. An experiment is performed. For a fiber loop containing a 210-m fiber, pulse replication and the FTTM of pulses with a PRI of 20 μs and pulse width of 1.20, 1.00, 0.85, and 0.65 μs are realized. The measurement errors are below ±12 MHz and ±5 MHz in the frequency range from 0.3-0.7, 0.9-1.3, and 1.5-1.9 GHz by using one pair of pulses and all the replicated pulses, respectively, with a certain chirp rate of 0.978 THz/s. In addition, the influence of the sweep frequency chirp rate and pulse width on the measurement error is also studied. For a certain pulse to be measured, to a certain extent, the faster the frequency sweep, the greater the frequency measurement error. For a specific sweep frequency chirp rate, the measurement error is almost unaffected by the pulse width to be measured.

## 2. System and principle

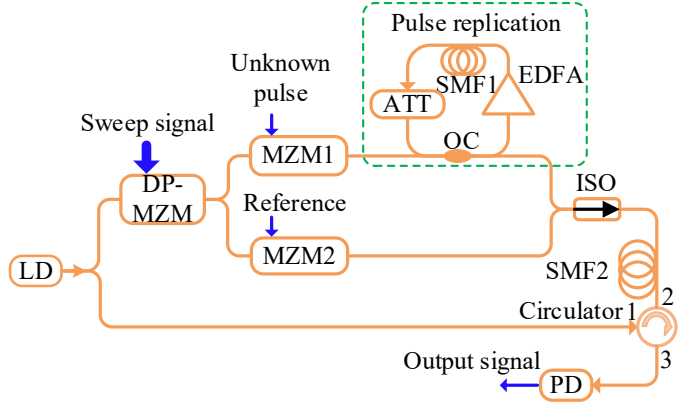

Fig. 1. The schematic diagram of the proposed microwave pulse detection and frequency measurement system. LD, laser diode; DP-MZM, dual-parallel Mach-Zehnder modulator; MZM, Mach-Zehnder modulator; EDFA, erbium-doped fiber amplifier; OC, optical coupler; SMF, single-mode fiber; ATT, attenuator; ISO, isolator; PD, photodetector.

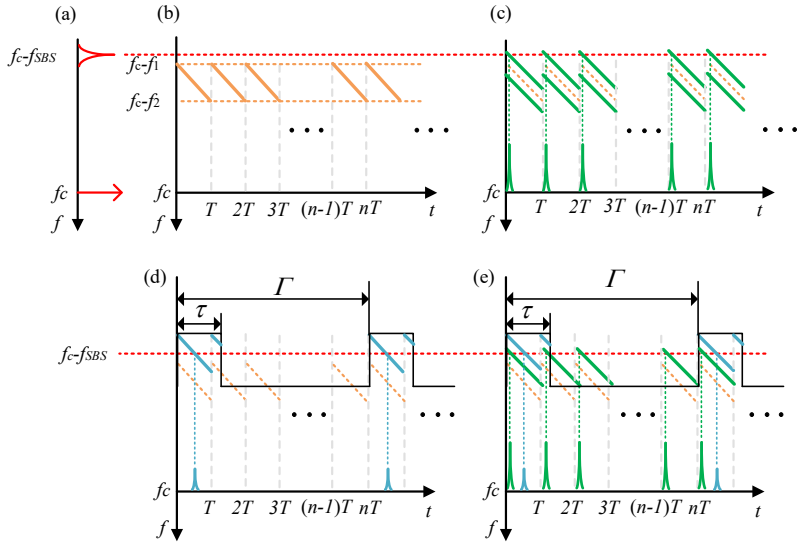

Fig. 2. Principle of operation. (a) The SBS gain with its frequency centered at $f_c$-$f_{SBS}$ generated by the optical carrier centered at $f_c$. (b) Time-frequency characteristics of the generated sweep optical carrier from the DP-MZM, where $T$ is the sweep period. (c) Time-frequency characteristics and the FTTM of the optical signal from MZM2. (d) Time-frequency characteristics and the FTTM of the optical signal from the optical fiber loop, where $\Gamma$ is the time delay of the loop and $\tau$ is the pulse width of the microwave pulse. (e) Time-frequency characteristics and the FTTM of the optical signal from the optical circulator. The positive sidebands that do not interact with the SBS gain during the scanning process are not given in (d) and (e).

Fig. 1 shows the schematic diagram of the proposed microwave pulse detection and frequency measurement system. A CW light wave generated from a laser diode (LD) is split into two branches. In the lower branch, the CW light wave used as the pump wave is injected into a spool of single-mode fiber (SMF2) via an optical circulator to induce the SBS effect, which will generate an SBS gain with its frequency centered at $f_c$-$f_{SBS}$ as shown in Fig. 2(a). In the upper branch, the CW light wave is CS-LSSB modulated at a DP-MZM by an electrical sweep signal with a period of $T$, a negative chirp rate of $k$, and a bandwidth ranging from $f_1$ to $f_2$ to generate a periodic sweep optical carrier, with its time-frequency characteristic shown in Fig. 2(b). Subsequently, the periodic sweep optical carrier is also split into two branches. In one branch, it is modulated at a null-biased Mach-Zehnder modulator (MZM2) by a CW single-frequency reference $f_r$, which aims to generate a reference optical signal, with its time-frequency characteristic and the FTTM shown in Fig. 2(c). Because MZM2 is null-biased, the reference optical signal is a CS-DSB signal. As can be seen from Fig. 2(c), the generated negative sidebands of the CW single-frequency reference can be amplified by the SBS gain at the initial time in every period during the scanning process, whereas the positive sidebands do not interact with the SBS gain. Therefore, low-frequency pulses with a PRI equal to the sweep period $T$ are generated in the time domain, which can be named as the reference pluses shown as the green pulses in Fig. 2(c). In the other branch, the periodic sweep optical carrier is modulated at a second null-biased MZM (MZM1) by the unknown microwave pulse with a pulse width of $\tau$ and a carrier frequency of $f$. To better capture and further identify the unknown pulse, the output of the MZM1 is injected into an optical fiber loop, which mainly consists of an erbium-doped fiber amplifier (EDFA), a spool of SMF (SMF1), and an optical attenuator (ATT), to achieve pulse replication. Note that, the loop delay $\Gamma$ should be larger than the microwave pulse width $\tau$. At the same time, the pulse width $\tau$ should be larger than the optical sweep period $T$. To make sure that the FTTM of the replicated pulse is synchronized with the original one, the period $T$ of the periodic sweep optical signal and the loop delay $\Gamma$ needs to satisfy the following condition:

$$\Gamma = nT, \qquad (1)$$

where $n$ is an integer. In Fig. 2(d), the SBS gain spectrum can be detected only once within $\tau$, because the pulse width $\tau$ is not much larger than the period $T$ of the sweep optical signal. If $\tau$ is much larger than $T$, the SBS gain spectrum can be detected multiple times within $\tau$. In the proposed system, the frequency of the microwave pulse can be measured, even if it can be detected by the SBS gain spectrum only once. Then, the optical signals from the optical fiber loop and MZM2 are combined and sent to SMF2 via an isolator as the probe wave, which will be detected by the SBS gain spectrum provided by the pump wave. As shown in Fig. 2(e), the reference pulses are observed in every period of the scanning process, whereas the pulses corresponding to the microwave pulse are only observed in the periods, during which the original optical pulse or the replicated pulse exists. To see it clear, the positive sidebands that do not interact with the SBS gain during the scanning process are not given in Fig. 2(d) and (e). As can be seen in Fig. 2(e), when two pulses in a sweep period are observed, the carrier frequency of the unknown microwave pulse can be determined by the time difference between the two low-frequency electrical pulses and the chirp rate $k$. Supposing the time difference between the two pulses is $\Delta T$, the carrier frequency of the unknown microwave pulse can be expressed as

$$f = f_r + |k| \Delta T. \qquad (2)$$

In the proposed scheme, the meaning of pulse replication in the optical domain is to capture and identify narrow microwave pulses more easily. Thanks to the pulse replication, multiple pairs of reference pulses and signal pulses are generated. The measurement of the carrier frequency is easier compared with the case with only one pair. In addition, in all usable periods where a pair of pulses are generated in a period, the time difference $\Delta T$ between the two pulses in a pair is theoretically the same, so we can obtain the frequency of the microwave pulse in any period. Furthermore, note that the time difference $\Delta\tau$ between the pulse replication system output and the MZM2 output reaching the SBS medium will lead to measurement errors. The carrier frequency of the unknown pulse can be expressed as

$$f = f_r + |k|(\Delta T + \Delta\tau). \tag{3}$$

As can be seen, there is an unwanted fixed error value $|k|\Delta\tau$ and it is necessary to match the two paths or calibrate the system to remove its influence in the experiments.

Note that the maximum value of $\Delta T$ is determined by the sweep period $T$ of the designed periodic sweep optical carrier, which indicates that the measurable frequency range is from $f_r$ to $f_r + |k|T$. Thus, the frequency measurement range is less than the bandwidth of the periodic sweep optical carrier. However, if $T$ is fixed, the chirp rate $|k|$ cannot be increased arbitrarily to increase the measurable range, which is due to that a too large chirp rate will result in poor frequency resolution and poor measurement accuracy. There is a trade-off between the measurement range and frequency resolution. The frequency measurement resolution can be improved by narrowing the SBS gain spectrum [22] or enhancing the Brillouin gain [23]. Furthermore, for a certain periodic sweep optical carrier with a period of $T$ and a chirp rate of $k$, the frequency measurement range can be changed by changing the reference frequency $f_r$. However, only changing the reference will bring about a problem: the measurement range becomes smaller. To ensure that the measurement range is not reduced, the reference frequency $f_r$ and the start frequency $f_1$ of the electrical sweep signal should meet the following condition:

$$f_r + f_1 = C, \tag{4}$$

where $C$ is a constant and just slightly larger than $f_{SBS}$. In this case, the generated negative sidebands from the CW single-frequency reference $f_r$ can be just amplified by the SBS gain at the initial time in every period during the scanning process.

## 3. Experiment results and discussion

*3.1. Experimental setup*

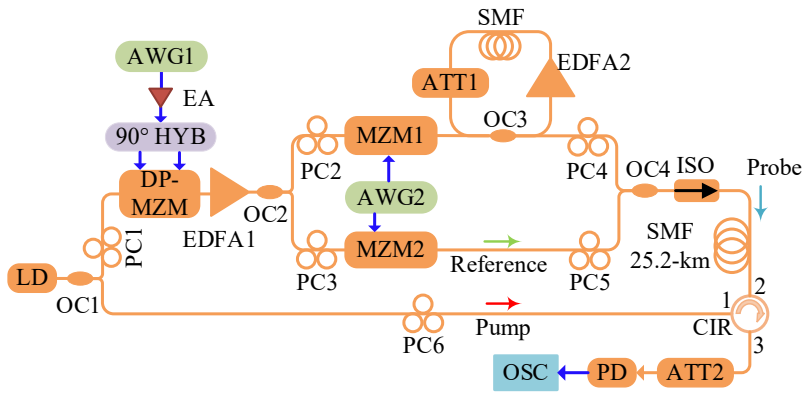

Fig. 3. Experimental setup of the proposed microwave pulse detection and frequency measurement system. LD, laser diode; OC, optical coupler; PC, polarization controller; DP-MZM, dual-parallel Mach-Zehnder modulator; MZM, Mach-Zehnder modulator; AWG, arbitrary waveform generator; 90° HYB, 90° electrical hybrid coupler; EDFA, erbium-doped fiber amplifier; EA, electrical amplifier; SMF, single-mode fiber; ATT, attenuator; ISO, isolator; CIR, circulator; PD, photodetector; OSC, oscilloscope.

An experiment based on the setup shown in Fig. 3 is performed to verify the proposed microwave pulse detection and frequency measurement system. A 15.5-dBm optical carrier centered at 1549.964 nm from the LD (ID Photonics, CoBriteDX1-1-C-H01-FA) is divided into two paths via OC1. In the upper path, the optical carrier is CS-LSSB modulated at the DP-MZM (Fujitsu, FTM7961EX) by a designed sweep electrical signal from AWG1 (Keysight M8195A) to generate a sweep optical carrier. The power of the sweep electrical signal is around -10 dBm and amplified by an electrical amplifier (EA, ALM/145-5023-293 5.85-14.5 GHz). Here, to achieve the CS-LSSB modulation, a 90° electrical hybrid coupler (90° HYB, Narda 4065 7.5–16 GHz) is used, and the two sub-MZMs of the DP-MZM are both null-biased and the main-MZM is biased to introduce a 90° phase shift. Subsequently, the output of the DP-MZM is divided into two paths via OC2 after being amplified by EDFA1 (Amonics, EDFA-PA-35-B). The output of EDFA 1 is set as 10.1 dBm. In one path, the sweep optical carrier is firstly CS-DSB modulated at the null-biased MZM1 (Fujitsu, FTM7938EZ) by the 10-dBm unknown microwave pulse from AWG2 (Keysight M8190A). Then, the output of MZM1 is injected into the pulse replication system via OC3, which is an optical fiber loop mainly consisting of EDFA2 (MAX-RAY PA-35-B), a spool of SMF, and an optical attenuator. The output of EDFA 2 is set as around 8.5 dBm. In the other path, the sweep optical carrier is CS-DSB modulated at the null-biased MZM2 (Fujitsu, FTM 7938EZ) by a 7-dBm fixed CW reference electrical signal from AWG2 to generate the reference optical signal. Then, the reference optical signal and the output of the pulse replication system are coupled together via OC4 and injected into the 25.2-km SMF through an optical isolator. Polarization controllers (PC1, PC2, and PC3) are used to optimize the light polarizations before the DP-MZM, MZM1, and MZM2, respectively. In the lower path, the optical carrier is used as the pump wave and launched into the 25.2-km SMF via an optical circulator, where it interacts with the counter-propagating wave from the upper branch. PC4, PC5, and PC6 are used to ensure the efficient stimulated Brillouin interaction. Then, the optical signal from the SMF is detected by a photodetector (PD, Nortel PP-10G) and monitored by an oscilloscope (OSC, R&S RTO2032). An electrical attenuator is inserted to prevent saturation of the optical power injected into the PD, and the optical power before the PD is about -23 dBm.

*3.2. Pulse replication based on an active fiber loop*

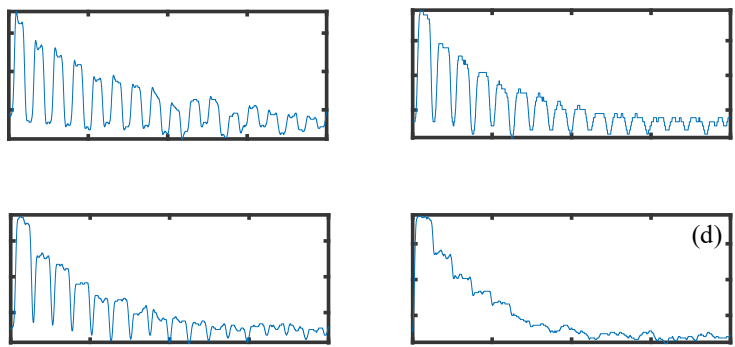

Fig. 4. The output pulse trains in one pulse period after pulse replication using a pulse with a PRI of 20 μs and a pulse width of (a) 0.65 μs, (b) 0.85 μs, (c) 1.00 μs, (d) 1.20 μs.

To capture and identify a pulse with a large PRI and further measure its carrier frequency much easily, the pulse needs to be firstly replicated. A pulse replication experiment based on the active fiber loop is firstly carried out. The length of the SMF in the fiber loop is chosen to be around 210 m, and the total time delay of the loop is measured to be 1.2264 μs. The unknown pulses with a PRI of 20 μs and a pulse width of 0.65, 0.85, 1.00, and 1.20 μs are chosen to demonstrate the pulse replication, respectively, with the waveforms of the output pulse trains in one pulse period after replication shown in Fig. 4. As expected, the pulse replication is successfully achieved and many pulses are replicated by controlling the gain of the loop. As can be seen from the results shown in Fig. 4(a) and (b), the time interval between the adjacent replicated pulses is consistent with the loop delay. However, as shown in Fig. 4(c) and (d), for the pulses with a pulse width of 1.00 and 1.20 μs, because the loop delay is slightly larger than the pulse width, there is no gap between the replicated pulses. In the fiber loop, the gain is slightly smaller than the loss, so the amplitudes of the replicated pulses gradually decrease. In fact, we can further increase the number of replicated pulses by balancing the gain and loss of the loop. In this case, an optical switch is needed to select the length of the replicated pulse to prevent the replicated pulse from meeting the next pulse injected into the fiber loop. In addition, when the single narrow pulse is replicated into a pulse train, further low-pass filtering will make the pulse train easier to be sampled, which makes it easier to detect the narrow pulses with large PRI.

*3.3. FTTM and frequency measurement*

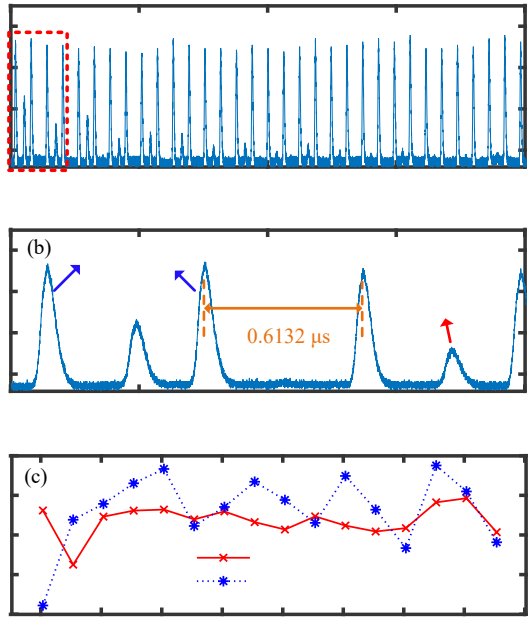

Fig. 5. (a) The waveform of the photocurrent from the PD when the microwave pulse has a carrier frequency of 0.525 GHz, a PRI of 20 μs, and a pulse width of 0.65 μs. (b) A zoom-in view of the waveform outlined in the red dotted box in (a). (c) The measurement errors of the pulses with different carrier frequencies ranging from 0.3 to 0.675 GHz with a frequency step of 25 MHz.

To measure the carrier frequency of the microwave pulse, the FTTM of the microwave pulse after pulse replication is implemented. The period, bandwidth, and center frequency of the sweep signal from AWG1 are set to 0.6132 μs, 0.6 GHz, and 10.4 GHz, respectively. The CW reference signal from AWG2 is fixed at 0.2 GHz unless otherwise specified in this paper. The microwave pulse with a carrier frequency of 0.525 GHz, a pulse width of 0.65 μs, and a PRI of 20 μs is chosen as the microwave pulse to be measured. Fig. 5(a) shows the waveform of the photocurrent from the PD. Fig. 5(b) shows the zoom-in view of the waveform outlined in the red dotted box shown in Fig. 5(a). As can be seen from Fig. 5(a) and (b), both the fixed CW reference and the carrier frequency of the microwave pulse have been mapped into the time domain. The high-amplitude reference pulses with a fixed time interval of 0.6132 μs are generated by the CW reference. The low-amplitude pulses, i.e. the signal pulses, are generated by the original microwave pulses and the replicated ones. Compared with the replicated pulses in Fig. 4(a), the low-amplitude pulses only exist in the time duration where a replicated pulse is generated. Because the period of the sweep signal is half the fiber loop delay and the pulse width is only a bit larger than the period of the sweep signal, as can be seen from Fig. 5(a) and (b), the low-frequency signal pulse appears once every other sweep period. By calculating the time difference between the reference pulse and the signal pulse, the carrier frequency of the microwave pulse can be obtained using Eq. (2). Fig. 5(c) shows the frequency measurement errors of the microwave pulses with carrier frequencies ranging from 0.3 to 0.675 GHz with a frequency step of 25 MHz. The errors are below ±10 MHz by using only one pair of pulses, with the results shown in blue stars in Fig. 5(c). Since multiple pairs of reference pulse and signal pulse are generated, the carrier frequency of the microwave pulse can further be obtained by averaging the results from multiple pairs of pulses, and the errors are below ±5 MHz as shown in the red crosses in Fig. 5(c), which indicates that the errors can be reduced by averaging multiple sets of FTTM results.

*3.3.1. The influence of the sweep frequency chirp rate*

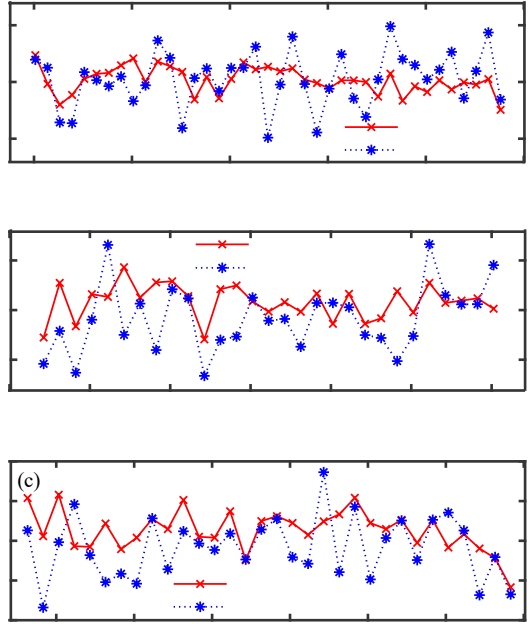

Fig.6. The frequency measurement errors of microwave pulses with different frequencies ranging from (a) 0.40 to 1.35 GHz with a frequency step of 25 MHz, (b) 0.6 to 2.0 GHz with a frequency step of 50 MHz, (c) 0.8 to 3.9 GHz with a frequency step of 100 MHz.

Then, the influence of the frequency sweep chirp rate on the pulse frequency measurement is studied. The period of the sweep signal from AWG1 is fixed at 0.6132 μs, while the sweep bandwidth is set to 1.4, 2.4, and 4.4 GHz, respectively. Accordingly, the center frequency of the sweep signal is set to 10, 9.5, and 8.5 GHz. The microwave pulse has a PRI of 20 μs and a pulse width of 0.65 μs. Fig. 6(a) shows the frequency measurement errors of microwave pulses with the carrier frequencies ranging from 0.4 to 1.35 GHz with a frequency step of 25 MHz, by using a sweep signal with a sweep bandwidth of 1.4 GHz. The blue stars show that the error is below ±25 MHz by using one pair of pulses, while the red crosses indicate that the error is reduced to below ±12 MHz by using multiple pairs of pulses. Fig. 6(b) shows the frequency measurement errors of the microwave pulses with the carrier frequencies ranging from 0.6 to 2 GHz with a frequency step of 50 MHz, by using a sweep signal with a sweep bandwidth of 2.4 GHz. The error is below ±35 MHz and below ±20 MHz by using one pair of pulses and by using multiple pairs of pulses, respectively. Fig. 6(c) shows the frequency errors of the pulses with the carrier frequencies ranging from 0.8 to 3.9 GHz with a frequency step of 100 MHz, by using a sweep signal with a sweep bandwidth of 4.4 GHz. The error is below ±90 MHz and below ±60 MHz by using one pair of pulses and by using multiple pairs of pulses, respectively. One can easily find that as the sweep bandwidth increases, i.e. the sweep chirp rate increases, the measurable frequency range becomes larger, but the measurement accuracy deteriorates, leading to a trade-off between the measurement range and accuracy.

Another way to change the chirp rate is to change the scanning period while the scanning bandwidth is fixed. From the results above, it is easy to understand that reducing the frequency sweeping chirp rate, that is, increasing the length of the frequency sweeping time, can improve the frequency measurement accuracy within a certain measurement range, at the cost of a longer time. Therefore, in real-world applications, it is necessary to select a suitable frequency sweeping chirp rate of $k$ and frequency sweeping period of $T$ according to different application requirements.

*3.3.2. The influence of the microwave pulse width*

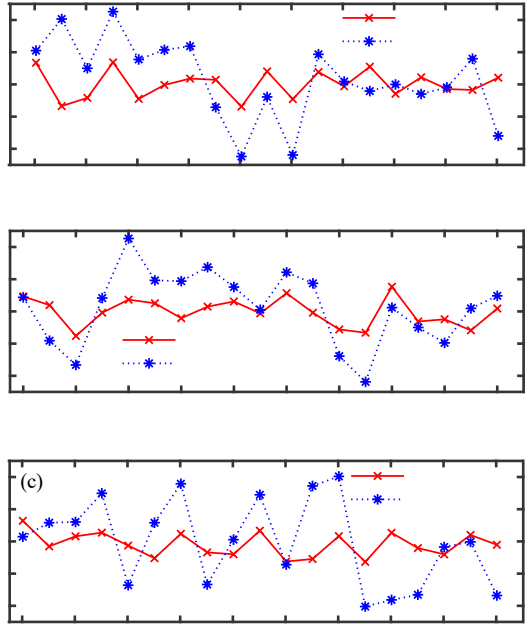

Fig.7. The measurement errors of the microwave pulses with a PRI of 20 μs and a pulse width of (a) 0.85, (b) 1.00, (c) 1.20 μs under a fixed sweep chirp rate of 2.283 THz/s.

The influence of the microwave pulse width on the frequency measurement is studied with a fixed sweep frequency chirp rate. The period, the bandwidth, and the center frequency of the sweep signal from AWG1 are set to 0.6132 μs, 1.4 GHz, and 10 GHz. The microwave pulses with a PRI of 20 μs and a pulse width of 0.85, 1.00, 1.20 μs are chosen as the signal to be measured. Fig. 7 shows the frequency measurement errors, which are all below ±25 MHz and below ±10 MHz by using one pair of pulses and multiple pairs of pulses, respectively, indicating that the microwave pulse width has almost no effect on the accuracy of the frequency measurement in the proposed scheme. The reason is that the width of the microwave pulses to be measured is greater than the frequency sweep period, and the fast sweep optical signal is the main factor that affects the SBS effect.

*3.3.3. Reconfigurable measurable frequency range*

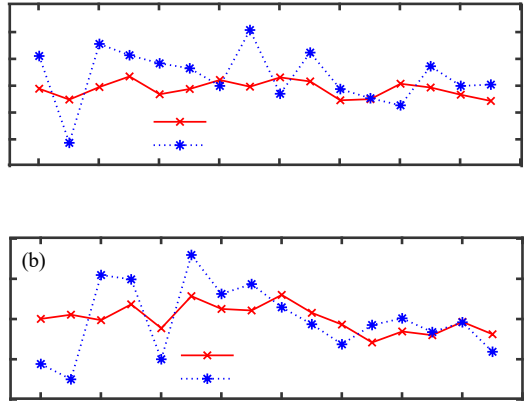

Fig.8. The measurement errors of the microwave pulses with different carrier frequencies ranging from (a) 0.9 to 1.3 GHz and (b) 1.5 to 1.9 GHz with a frequency step of 25 MHz. The PRI and width of the microwave pulses are 20 and 0.65 μs, while the sweep chirp rate is 0.978 THz/s.

Reconfigurable frequency measurement range is demonstrated when both the frequency sweeping chirp rate and the pulse width are fixed. As mentioned above, the frequency measurement range is less than the sweep bandwidth of the periodic sweep optical signal. In the experiment, to measure the microwave pulse with a pulse width of 0.65 μs and a PRI of 20 μs, the sweep bandwidth is set to 0.6 GHz. To change the frequency measurement range, based on Eq. (4), the frequency of the CW reference signal is changed to 0.8 and 1.4 GHz for the frequency measurement ranges of 0.9 to 1.3 GHz and 1.5 to 1.9 GHz, respectively, while the center frequency of the sweep optical signal is set to 9.8 and 9.2 GHz with a fixed sweep bandwidth of 0.6 GHz. Fig. 8(a) shows the frequency measurement errors of the pulses with carrier frequencies ranging from 0.9 to 1.3 GHz with a frequency step of 25 MHz. The errors are below ±12 MHz and below ±5 MHz by using one pair of pulses and using multiple pairs of pulses, respectively. Fig. 8(b) shows the frequency measurement errors of the pulses with the carrier frequencies ranging from 1.5 to 1.9 GHz with a frequency step of 25 MHz. The errors are below ±10 MHz and below ±5 MHz by using one pair of pulses and using multiple pairs of pulses, respectively. Compared with Fig. 5(c), the errors of the three measurement frequency bands show good consistency, which also indicates the reliability of the system to a certain extent.

## 4. Conclusion

In conclusion, a photonics-assisted pulse detection and frequency measurement system has been proposed and experimentally demonstrated based on pulse replication and SBS-based FTTM. The key significance of the work is that a pulsed microwave signal is captured and measured with the help of pulse replication and SBS-based FTTM for the first time. Pulse replication makes it possible to detect microwave pulses with large PRI, whereas FTTM maps the microwave pulses to low-frequency electrical pulses for pulse frequency measurement. Furthermore, the pulse frequency measurement accuracy can also be improved by using multiple replicated microwave pulses. An experiment is performed. For a fiber loop containing a 210-m fiber, pulse replication and the FTTM of pulses with a PRI of 20 μs and pulse width of 1.20, 1.00, 0.85, and 0.65 μs are realized. Under a certain sweep frequency chirp rate of 0.978 THz/s, the measurement errors are below ±12 MHz and below ±5 MHz within a frequency range from 0.3-0.7, 0.9-1.3, and 1.5-1.9 GHz by using one pair of pulses and multiple pairs of pulses. In addition, the influence of sweep frequency chirp rate and pulse width on measurement error has also been studied. For a certain pulse to be measured, to a certain extent, the faster the frequency sweep, the greater the frequency measurement error. For a specific sweep frequency chirp rate, the measurement error is almost unaffected by the pulse width to be measured. This work provides an optical solution for the detection and identification of microwave pulses and is expected to play an important role with the help of integrated optoelectronics technology.

## Acknowledgements

This work was supported in part by the Natural Science Foundation of Shanghai under Grant 20ZR1416100, in part by the National Natural Science Foundation of China under Grant 61971193, in part by the Open Fund of State Key Laboratory of Advanced Optical Communication Systems and Networks, Peking University, China, under Grant 2020GZKF005, and in part by the Science and Technology Commission of Shanghai Municipality under Grant 18DZ2270800.